\documentclass{moriond}

\def\be{\begin{equation}}
\def\ee{\end{equation}}
\def\bea{\begin{eqnarray}}
\def\eea{\end{eqnarray}}

\newcommand{\Photo}{}

\begin{document}
\vspace*{4cm}
\title{Cosmology of Massive Gravity and its Extensions}

\author{ Kurt Hinterbichler\vspace{0.5cm}}

\address{Department of Physics, Case Western Reserve University, \\
10900 Euclid Ave, Cleveland, OH 44106, USA}

\maketitle\abstracts{
We review the status of the theory of dRGT massive gravity, and some of its extensions, as applied to cosmology and the cosmological constant problem.}

\section{Introduction}

The cosmological constant problem \cite{Weinberg:1988cp,Padilla:2015aaa} is a theoretical loose end in the description of the universe provided by the $\Lambda$CDM model of cosmology.
One can categorize possible solutions on a spectrum ranging from conservative to radical.  The multiverse/anthropic explanation sits on the conservative side, because in this case there is no new physics directly invoked to explain the acceleration; there is nothing to dark energy beyond a pure cosmological constant and its small value is explained anthropically via a selection effect within a dynamically populated multiverse of many possible values.  The lack of any widely agreed upon compelling alternatives has driven many theorists to this conclusion.  On the radical side, we have proposals which seek to change the rules of the game; perhaps effective field theory or quantum mechanics fundamentally fails at cosmological distances and something else is going on.

In between the two extremes, we have proposals which seek to address the problem by introducing new light degrees of freedom (they must be light so as to affect cosmology on large distances) without changing the usual rules of effective field theory.   If we wish to maintain standard assumptions such as Lorentz invariance and locality,\footnote{These are assumptions which are well tested, but there is always the possibility that they may be violated cosmologically in the dark sector, and there are many interesting proposals along these directions, e.g. \cite{Dubovsky:2004sg,Berezhiani:2007zf,Rubakov:2008nh,Comelli:2013txa,Jaccard:2013gla,DeFelice:2015hla}.} then there is a list of possible representations (classified by Wigner and others) and we need only look at this list and decide what to add.

One representation in particular raises some questions -- the massive spin-2 representation.  General relativity describes the graviton as a massless spin-2, carrying two physical degrees of freedom.  Is there any reason why the graviton could not live in the massive representation instead?  What would physics look like if it did \footnote{The question of whether the graviton has a small mass falls into the class of natural ``what if" questions which are important to ask in their own right, regardless of whether they may help solve some outstanding problem or be tested observationally.  If a program based on such a question fails for whatever theoretical or phenomenological reason, it is not a write-down, because we learn something about why the universe is as it is -- one of the best way to learn about what is, is by probing what could have been.
}?  

There is a heuristic reason why we might think a small mass could help with the cosmic acceleration.  The potential generated by a massless particle is the familiar Coulomb potential $\sim {1\over r}$, but a massive particle instead generates a Yukawa potential $\sim {1\over r}e^{-mr}$ which has an exponential suppression kicking in at the length scale $\sim {1\over m}$.  If we imagine that this scale is very large, on the order of the Hubble scale today, $m\sim H$, then the force of a massive graviton would be weakened at these large scales and perhaps this could say something about the acceleration.  Of course, this is a just a heuristic thought gleaned from the linear theory, and one should not try to take it too far.  A proper treatment requires finding a fully non-linear theory and studying its cosmology. 

\section{Non-linear massive spin-2}

A massive spin-2 has five degrees of freedom.  The three degrees of freedom beyond the two of the massless graviton are the extra light modes we hope will be useful in explaining the current acceleration of the universe.  The reason we may hope that this gives a better explanation than the cosmological constant is a symmetry reason: GR has a diffeomorphism symmetry which is broken in the massive theory.  As the mass goes to zero, this symmetry is restored, so any quantum corrections to the mass (from e.g. heavy particles) should be proportional to the mass itself \footnote{Diffeomorphism symmetry is a gauge symmetry, so the argument is more subtle than this, but the same conclusion bears out \cite{deRham:2012ew,deRham:2013qqa}, for essentially the same reasons the mass of the W and Z bosons can be naturally small compared with the scale of electroweak symmetry breaking.}, and thus a small mass is technically natural.

It was long thought that a non-linear theory of a pure massive spin-2 could not be written down.  This is primarily due to the Boulware-Deser ghost problem \cite{Boulware:1973my}: if one adds generic self-interaction terms to the linear \cite{Fierz:1939ix} massive spin-2 theory, one finds that the non-linear theory propagates six physical degrees of freedom, as opposed to the desired 5 physical degrees of freedom.  Around flat space, this extra degree of freedom is a non-issue: the theory is an effective field theory with a strong coupling scale $\Lambda_5\sim (m^4 M_P)^{1/5}$, perturbatively the theory still has five degrees of freedom, and the ghost does not appear until one goes beyond this cutoff so one cannot say it is in the theory.  The real problem is that one cannot say that any interesting non-linear solutions are in the theory \cite{ArkaniHamed:2002sp,Deffayet:2005ys,Creminelli:2005qk}.
 If we attempt to find non-linear solutions (such as cosmological ones) and build an effective theory around them, the extra degree of freedom will generally appear as an additional scalar mode which has a wrong sign kinetic term and signals an instability in the background.  
 
 However, just because generic interactions have this issue, it does not mean that all interactions do, and in fact the renewed interest in massive gravity theories over the past few years is primarily due to the development of a theory with interactions that give a higher strong coupling scale and avoid this ghost issue.  The theory was found by using modern effective field theory techniques to find interactions order by order \cite{deRham:2010ik}, and then re-summing the result \cite{deRham:2010kj}.  It is known as dRGT theory, after de Rham, Gabadadze and Tolley (see \cite{Hinterbichler:2011tt,deRham:2014zqa} for reviews).  The action consists of the Einstein-Hilbert term plus a specific two parameter family of zero derivative potential terms,
 \be S={M_P^{2}\over 2}\int d^4x\, \sqrt{-g}\left[R[g]-{m^2\over 4}\sum_{n=0}^3\beta_n S_n\left(\sqrt{g^{-1}\eta}\right)\right]\, . \label{massivegac}\ee   
Here $M_P$ is the usual Planck mass, $m$ is the graviton mass, $\eta_{\mu\nu}$ is a fixed fiducial Minkowski metric upon which the theory is built, and $g_{\mu\nu}$ is the dynamical metric.  $S_n$ are the symmetric polynomials of the matrix square root $M^\mu_{\ \nu}\equiv\sqrt{g^{\mu\lambda}\eta_{\lambda\nu}}$, $S_n(M) =M^{ [\mu_1}_{\ \mu_1}\cdots M^{ \mu_n]}_{\ \mu_n}$.  Of the 4 dimensionless coefficients $\beta_0,\cdots,\beta_3$, one combination should be tuned to allow a Minkowski solution and another combination is redundant with the graviton mass, leaving a two parameter family of interaction terms.  The mass terms can be thought of as bi-metric generalizations of the cosmological constant, which can be given a geometric interpretation \cite{Hinterbichler:2012cn,Gabadadze:2013ria,Goon:2014paa}.
There are by now many proofs that the theory is ghost free \cite{Hassan:2011hr,Hassan:2011ea,Mirbabayi:2011aa,Hassan:2012qv}.
 
 This theory is the unique way to add zero-derivative interaction terms to the Einstein-Hilbert action and maintain only 5 degrees of freedom (though there are other ways without using the Einstein-Hilbert term \cite{Folkerts:2011ev,Hinterbichler:2013eza}).  It is an effective field theory with a strong coupling scale $\Lambda_3\sim (m^2M_P)^{1/3}$, which is the highest possible strong coupling scale around Minkowski space for any theory of pure massive spin-2 build by adding zero-derivative interaction terms to Einstein-Hilbert \cite{Schwartz:2003vj,Cheung:2016yqr}.  Beyond this strong coupling scale, one finds unphysical effects such as superluminal propagation and breakdown of the Cauchy problem \cite{Deser:2012qx,Brito:2014ifa,Motloch:2015gta} in the classical theory.  It is possible to dynamically reach these regions, so the main outstanding theoretical issue is UV completion: some new physics or strong coupling effects must kick in to complete the theory and determine what happens beyond this strong coupling scale.

\section{Cosmological solutions of dRGT gravity} 

The general intuition is that massive gravity is a small deformation of GR, so the physics should be GR-like to the extent that the graviton mass is small.  The number of degrees of freedom, however, is not continuous, and the way that physical continuity is achieved is through non-linearities which screen the extra degrees of freedom and cause them to decouple -- the so-called Vainshtein mechanism \footnote{Non-linearities are essential for the operation of the Vainshtein mechanism; it does not work in the purely linear theory, which is why we have the van Dam-Veltman-Zakharov (vDVZ) discontinuity \cite{vanDam:1970vg,Zakharov:1970cc}.} \cite{Vainshtein:1972sx,Deffayet:2001uk,Babichev:2009jt,Babichev:2010jd} (see \cite{Babichev:2013usa} for a review).  Around a source of mass $M$, there is a new distance scale $r_\ast\sim \left(M\over M_P\right)^{1/3}{1\over \Lambda_3}$, known as the Vainshtein radius, at which the non-linearities of the extra degrees of freedom start to become important (the non-linearities of GR do not become important until the much smaller Schwarzschild radius $r_s\sim {M\over M_P^2}$).  Inside the Vainshtein radius, the effects of the extra degrees of freedom are screened and physics looks like that of GR, whereas outside the Vainshtein radius the extra degrees of freedom are active and can give ${\cal O}(1)$ deviations from GR.  Traditional tests of gravity, such as solar system tests, are all passed because these tests are performed well within the Vainshtein radius of the relevant gravitating body.

The intuition for cosmology is the following.  Define an energy density associated with the graviton mass scale, $\rho_c\sim M_P^2m^2$.  Consider some sphere of matter of radius $R$ and density $\rho$.  The Vainshtein radius of this sphere of matter is $r_\ast\sim \left(\rho\over \rho_c\right)^{1/3}R$.  If $\rho\gg \rho_c$, then $r_\ast \gg R$ and the sphere of matter lies 
within its own Vainshtein radius and behaves as in GR.  If $\rho\ll \rho_c$, then $r_\ast\ll R$ and the sphere of matter is larger than its own Vainshtein radius and experiences ${\cal O}(1)$ deviations from GR.  Consider the very early universe, so that $H\gg m$ and $\rho \gg \rho_c$.  Each Hubble patch of radius ${1/H}$ is well within its own Vainshtein radius and the universe behaves as in GR.   In particular, all the physics of inflation, the CMB, etc. are unaffected.  Now consider the far future universe, where $H\ll m$ and $\rho \ll \rho_c$.  In this case each Hubble volume now lies outside its own Vainshtein radius so the behavior will be drastically different from GR.  
The crossover time occurs when $H\sim m$ and $\rho\sim \rho_c$.  If we want the graviton to be explaining the current acceleration, the crossover should be happening about now.

When one tries to realize this quantitatively by solving the equations of motion of the action Eq.\ref{massivegac} 
there are some immediate obstacles (see \cite{DeFelice:2013bxa} for a review).  The equations of motion for a standard flat FRW ansatz imply that the only solution for the scale factor $a(t)$ is a constant one: ${d\over dt}a=0$, so there are no flat FRW solutions other than Minkowski space (regardless of the presence of matter) \cite{D'Amico:2011jj}.  Closed FRW solutions also do not exist, but open FRW solutions can be found, and they come in two branches: the normal branch and the self-accelerating branch.  The normal branch is just Minkowski in a different slicing, but the self-accelerating branch is particularly interesting because the cosmology displays an acceleration with a scale $H\sim m$ set by the graviton mass, even in the absence of a bare cosmological constant \cite{deRham:2010tw,Koyama:2011xz,Nieuwenhuizen:2011sq,Chamseddine:2011bu,D'Amico:2011jj,Gumrukcuoglu:2011ew,Berezhiani:2011mt}.  However, upon studying perturbations, it is found that the scalar and vector modes around the self-accelerating solution have vanishing kinetic terms \cite{Gumrukcuoglu:2011zh,DeFelice:2012mx}.  This does not mean that the solutions are inconsistent or ruled out, but rather they are infinitely strongly coupled, so standard perturbative techniques cannot be used.  

There are other cosmological solutions that exhibit self-accelerating behavior with well-behaved perturbations, but which give up on the assumption of isotropy and have a preferred direction \cite{Gumrukcuoglu:2012aa,DeFelice:2013awa}.  It is these solutions which are expected to be the relevant physical cosmological solutions.  It seems that the massive theory prefers to be anisotropic, with the anisotropy becoming important only at late times (an interesting thing to consider given that observational constraints on anisotropies at the current Hubble scale may soon be tightened).
At very late times, when $H\ll m$, we can imagine a universe composed of domains of size $\sim {1\over m}$, each of which has its own randomly oriented preferred direction, but which taken together are isotropic on very large scales $\gg {1\over m}$.

Studying the detailed phenomenology of these anisotropic solutions is difficult because almost all the standard analysis tools and codes used in cosmology assume homogeneity and isotropy from the outset.  Because of this, little phenomenological work in this direction has been done so far, though progress can be made in certain limits \cite{deRham:2010tw,Spolyar:2013maa}.

\section{Extensions of Massive Gravity}

Most work on cosmology in massive theories has instead proceeded by attempting to keep the assumptions of homogeneity and isotropy and instead generalizing the theory so that it allows cosmological solutions.  These approaches all involve adding more degrees of freedom, while retaining the essential feature of having a massive spin-2 degree of freedom.

One early attempt was to introduce a new scalar field $\phi$ and promote the mass term (and/or couplings) to be functions of the scalar, $m\rightarrow m(\phi)$, known as {mass varying massive gravity} \cite{D'Amico:2011jj,Huang:2012pe}.  In this case, flat cosmological solutions with stable perturbations can be found \cite{Gumrukcuoglu:2013nza}, though it is not quite surprising or convincing because there are now arbitrary free functions in the Lagrangian which can be chosen so that this is the case.

One can introduce a scalar in a way that does not also introduce free functions by demanding symmetry under a kind of scaling symmetry, resulting in a theory known as {quasi-dilaton} \cite{D'Amico:2012zv}.  This theory has flat self-accelerating FRW solutions, but their perturbations are not completely stable and have other unusual features\cite{Gumrukcuoglu:2013nza,D'Amico:2013kya,Anselmi:2015zva}. 
There is an extension of the quasi-dilaton obtained by adding derivative operators consistent with the scale symmetry \cite{DeFelice:2013tsa}.  This extension solves the issue with the unstable perturbations and appears to be the first example of a theory with a stable self-accelerating cosmological solution \cite{Gabadadze:2014kaa,Kahniashvili:2014wua,Motohashi:2014una,Heisenberg:2015voa}.

Another possibility is to couple galileon-like scalars to the theory in a way which respects the galileon shift symmetry \cite{Gabadadze:2012tr} and preserves the absence of ghosts \cite{Andrews:2013ora}.  Geometrically, this is a natural thing to consider; the massive theory can be thought of as a sigma model from four-dimensional space time to some flat four-dimensional target space, and one gets the galileons by simply extending the dimensionality of the target space (coupling to gravity the brane constructions of \cite{deRham:2010eu,Hinterbichler:2010xn,Goon:2011uw,Goon:2011qf,Burrage:2011bt}).  The cosmological solutions and perturbations of this theory turn out to be very similar to those of pure dRGT theory; there are no flat FRW solutions and self-accelerating solutions have perturbations with vanishing kinetic terms \cite{Andrews:2013uca,Goon:2014ywa}.  An open question is whether stable non-isotropic solutions exist. 

Massive gravity is formulated with a fixed reference metric, and this metric can be made arbitrary without re-introducing a ghost \cite{Hassan:2011tf}.  One can attempt to make this reference metric cosmological, but this runs into tensions between keeping perturbations stable and maintaining the Vainshtein mechanism \cite{Fasiello:2012rw}.  A natural extension is to promote the background metric to a dynamical metric, adding a second Einstein-Hilbert term for it \cite{Hassan:2011zd}.  This preserves the ghost-freedom and strong coupling scale of the theory, resulting in an effective field theory describing a massless graviton and a massive graviton interacting with each other, for a total of seven degrees of freedom.  This is bi-metric gravity, and it is the extension which has received the most interest and analysis thus far (see \cite{Schmidt-May:2015vnx,Bull:2015stt,Solomon:2015hja} for reviews and further references).  

The study of cosmological solutions and their phenomenology in bi-gravity has been an active area recently \footnote{With two metrics, there is also the possibility for having non-trivial matter couplings, and there has been plenty of work exploring this possibility as well \cite{Hassan:2012wr,Akrami:2013ffa,deRham:2014naa,Noller:2014sta,Hassan:2014gta,deRham:2014fha,Soloviev:2014eea,Akrami:2014lja,Solomon:2014iwa,Gao:2014xaa,Yamashita:2014fga,Gumrukcuoglu:2014xba,Heisenberg:2014rka,Enander:2014xga,Schmidt-May:2014xla,Gumrukcuoglu:2015nua,Comelli:2015pua,Hinterbichler:2015yaa,Heisenberg:2016spl,Melville:2015dba,Lagos:2015sya,Heisenberg:2015iqa,Huang:2015yga}.} \cite{vonStrauss:2011mq,Volkov:2011an,Comelli:2011zm,Volkov:2012cf,Comelli:2012db,Berg:2012kn,Akrami:2012vf,Volkov:2013roa,Akrami:2013pna,Solomon:2014dua,Konnig:2014xva,Hassan:2014vja,DeFelice:2014nja,Cusin:2014psa,Enander:2015vja,Cusin:2015tmf,Nersisyan:2015oha,Konnig:2015lfa,Amendola:2015tua,Aoki:2015xqa,Akrami:2015qga}.  Here we give only a brief summary of the current status.  The cosmological solutions come in two branches.  First there is the non-dynamical (or algebraic) branch, which is the cousin of the self-accelerating branch of the pure massive theory.  On this branch, some of the perturbations have vanishing kinetic terms and so it is infinitely strongly coupled and not much can be done.  Then there is the dynamical branch, which has several sub-branches.  There is an ``infinite" sub-branch where the ratio of scale factors of the two metrics becomes infinitely large at early times and decreases at later times, but in general the perturbations on this sub-branch are ghostly in both the scalar and tensor sectors.  There is an ``exotic" sub-branch which describes bouncing cosmologies and other unconventional solutions (about which perturbations seem to generically have pathologies).  Then there is the ``finite" sub-branch, which has received the most attention so far.  In this sub-branch, there are self-accelerating solutions about which perturbations are healthy, with the exception of a scalar gradient instability which occurs at early times.   By tuning (in a technically natural way \cite{deRham:2012ew}) the parameters of the theory, this instability can be pushed to unobservably early times, but then the cosmological signatures become indistinguishable from $\Lambda$CDM.


\section{Conclusions}\label{subsec:prod}

The trend seems to be that if a model does not have a limit that looks like GR, it gets ruled out, and if it does have a limit that looks like GR, it gets pushed into this limit and becomes observationally indistinguishable \footnote{By indistinguishable we mean indistinguishable given foreseeable technology.  It is a different theory than GR, so it is of course in principle always distinguishable.} from GR.   Nevertheless, the fact that some of these models can explain acceleration without a cosmological constant is progress, even if it cannot at present be observationally distinguished from GR.  It provides a different interpretation of the data and of the cosmological constant problem.

So does any of this solve the cosmological constant problem?
Most would say that the answer, at present, is {\it no}, because it does not address the hard problem of why a large cosmological constant (from e.g. matter loops, phase transitions, bare CC, etc...) does not gravitate.   For this more is needed.  For instance, the idea of 
degravitation \cite{Dvali:2002fz,Dvali:2002pe,ArkaniHamed:2002fu,Dvali:2007kt}, that a condensate of massive gravitons can act as a high pass filter that effectively screens the effects of a cosmological constant.   This seems to works at the linear level, but it has not been convincingly demonstrated in a complete and fully non-linear theory, and there are general arguments against it \cite{Polchinski:2006gy,Bousso:2007gp}.  Though screening solutions can often be found (solutions in which the curvature is small despite a large bare cosmological constant), typical problems are that they do not possess a Vainshtein mechanism and thus run afoul of solar system tests \cite{deRham:2010tw}, or are not robust in the presence of matter.  There are, however, interesting proposals that make use of the essential structure of massive gravity \cite{Gabadadze:2014rwa,Gabadadze:2015goa}, and there is the lingering possibility that a novel kind of gauge symmetry present in the so-called ``partially massless" graviton may provide a symmetry that fixes the cosmological constant to the (already technically natural) small graviton mass, thereby providing a mechanism to keep the CC naturally small \cite{deRham:2013wv}.

Given the more modest goal of providing an alternative mechanism to explain the cosmic acceleration, without a cosmological constant (that is, assuming some unknown mechanism sets the expected large cosmological constant to zero), then some of these theories offer a solution.  This does not address the old cosmological constant problem of why the vacuum doesn't gravitate, but it reduces the problem to the form it took before the discovery of the cosmic acceleration.

%

\end{document}